\documentclass[11pt]{article}
\usepackage[sc]{mathpazo} 
\usepackage{fullpage}
\usepackage[authoryear,sectionbib,sort]{natbib}
\usepackage[utf8]{inputenc}
\usepackage{lineno}
\usepackage{titlesec}
\usepackage{setspace}
\usepackage{authblk}
\usepackage{natbib}
\usepackage{dsfont}
\usepackage{graphicx}
\usepackage{amsmath}
\usepackage{color, soul}
\usepackage{todonotes}
\usepackage{hyperref}
\usepackage{mathtools}
\titleformat{\section}[block]{\Large\bfseries\filcenter}{\thesection}{1em}{}
\titleformat{\subsection}[block]{\Large\itshape\filcenter}{\thesubsection}{1em}{}
\titleformat{\subsubsection}[block]{\large\itshape}{\thesubsubsection}{1em}{}
\titleformat{\paragraph}[runin]{\itshape}{\theparagraph}{1em}{}[. ]

\title{Simulating how animals learn: a new modelling framework applied to the process of optimal foraging}

\author{Peter R. Thompson$^{1,\ast}$ \\ 
M\'elodie Kunegel-Lion$^{2}$ \\ 
Mark A. Lewis$^{1, 3, 4, 5}$}

\date{}

\begin{document}

\maketitle

\noindent{} 1. Department of Biological Sciences, University of Alberta, Edmonton, AB, Canada;

\noindent{} 2. Canadian Forest Service, Northern Forestry Centre, Edmonton, AB, Canada;

\noindent{} 3. Department of Mathematical and Statistical Sciences, University of Alberta, Edmonton, AB, Canada.

\noindent{} 4. Department of Biology, University of Victoria, Victoria, BC, Canada.

\noindent{} 5. Department of Mathematics and Statistics, University of Victoria, Victoria, BC, Canada.

\noindent{} $\ast$ Corresponding author; e-mail: pt1@ualberta.ca.

\bigskip


\bigskip

\textit{Manuscript type}: Article. 

\bigskip

\noindent{\footnotesize Prepared using the suggested \LaTeX{} template for \textit{Am.\ Nat.}}

\linenumbers{}
\modulolinenumbers[3]

\newpage{}

\section*{Abstract}

Animal learning has interested ecologists and psychologists for over a century. Mathematical models that explain how animals store and recall information have gained attention recently. Central to this work is statistical decision theory (SDT), which relates information uptake in animals to Bayesian inference. SDT effectively explains many learning tasks in animals, but extending this theory to predict how animals will learn in changing environments still poses a challenge for ecologists. We addressed this shortcoming with a novel implementation of Bayesian Markov Chain Monte Carlo (MCMC) sampling to simulate how animals sample environmental information and learn as a result. We applied our framework to an individual-based model simulating complex foraging tasks encountered by wild animals. Simulated ``animals" learned behavioral strategies that optimized foraging returns simply by following the principles of an MCMC sampler. In these simulations, behavioral plasticity was most conducive to efficient foraging in unpredictable and uncertain environments. Our model suggests that animals prioritize highly concentrated resources even when these resources are less available overall, in line with existing knowledge on optimal foraging and ideal free distribution theory. Our innovative computational modelling framework can be applied more widely to simulate the learning of many other tasks in animals and humans.

\newpage{}

\section*{Introduction}


Animals do not know everything about the environments they live in \citep{Fagan2013}, and even if they did, the world is becoming more unpredictable every day \citep{IPCC2021}. While evolutionary adaptations are typically too slow to match these changes \citep{Bell2008, Chevin2010, Merila2014}, many animals can exhibit multiple behavioral responses to a changing environment without modifying their genetic code in a phenomenon known as behavioral plasticity \citep{DeWitt1998, Schmidt2010, Snell2013, Wong2015}. Examples of behavioral plasticity range from temporal adjustments in the phenology of frogs in the temperate forests of the eastern United States \citep{Gibbs2001} to the settlement of urban areas by birds in Europe \citep{Moller2009}. The ability to incorporate external information into a revised behavioral strategy may confer a fitness benefit to animals living in uncertain environments \citep{Parrish2000, Donaldson2008}, but the conditions in which behavioral plasticity is adaptive are not well-understood \citep{Wong2015}. Most forms of behavioral plasticity involve learning \citep{Snell2013}, which has a rich theoretical background \citep{Pearce2008} that could provide important context to the problem.

Our understanding of how animals learn is largely derived from laboratory studies of simple tasks \citep{Pearce2008}, which elucidate important cognitive mechanisms for learning but do not particularly resemble the natural world. This rich field of study can be traced back to Pavlov's work on conditioning and associative learning \citep{Pavlov1927, Harris2020}, which spawned theoretical and experimental work assessing the formation and extinction of these associative relationships, along with an animal's ability to categorize stimuli into different groups \citep{Spence1936, Rescorla1972, Pearce1987, Katz2006}. As food is often used as a positive reinforcer for animals \citep{Pavlov1927}, it follows logically that ``foraging" tasks can effectively display how animals learn to prioritize different food resources based on their relative reward \citep{Krebs1978, Lea2012}. Many of these conclusions draw heavily from optimal foraging theory \citep{Charnov1976}, generating a connection between cognitive and spatial ecology. When the proper data are available, animal movement and foraging processes are among the easiest methods for characterizing memory and learning in wild animals \citep{Fagan2013, Lewis2021}. The mechanistic clarity of laboratory experiments and the realism of animal movement models are difficult to combine into one analysis, but individual-based simulation modelling may be an effective tool for generating realistic patterns with clear mechanistic origins \citep{Tang2010, DeAngelis2019, Murphy2020}.

Cognitive psychologists and ecologists have identified a striking resemblance between learning and Bayesian inference. This is most clear when couched in terms of statistical decision theory (SDT; \citealp{Berger1985}). Broadly speaking, SDT is a mathematical framework describing the optimal way animals or humans should make decisions according to learned information \citep{Dall2005, Dayan2008, Schmidt2010}. A key component of SDT is the use of Bayes's theorem to represent how prior knowledge is updated through learning to produce a refined, posterior distribution of belief \citep{Berger1985}. Bayes's theorem and its key principles have been used to explain many learning processes \citep{Griffiths2001, McNamara2006}, including Pavlovian conditioning \citep{Courville2006}, mate choice \citep{Luttbeg1996, Castellano2012}, and optimal foraging \citep{Green1980, Green2006, Valone2006}. The application of SDT to optimal foraging problems has inspired the term ``Bayesian foraging", which describes how animals update their foraging preferences in a decision-theoretic manner \citep{Green1980, Valone2006}. Most of this work has focused on small-scale foraging tasks, but in reality, foraging is a very complex process influenced by many cognitive cues \citep{Fagan2013}. Extending SDT to a model that wholly encompasses animal movement and foraging will produce results that are more realistic and applicable to vulnerable wildlife populations.

Bayesian Markov Chain Monte Carlo (MCMC) sampling is a simple algorithm that we can use to simulate how animals learn. MCMC sampling uses a stochastic approach to calculate the posterior distribution of a set of parameters based on prior distributions and data supplied by the user \citep{Raftery1992}. When applied to learning, these parameters represent biological qualities of an animal, and the data represent information collected by animals through empirical experience. The structure of the prior and posterior distributions reflects the relative ``belief" an animal possesses in a certain behavioral strategy (i.e., combination of parameters) before and after incorporating ``data", respectively. The data enforce revised posterior belief in certain behavioral strategies through an objective function, which depends on the parameters and may also be stochastic. While the objective function in a MCMC sampling procedure is typically a probability distribution function (or likelihood function) of some sort, it does not need to be continuous nor does it need to integrate to 1 over the sampled domain. Instead of using MCMC to find the global optimum of a likelihood function, we can use it to identify behavioral strategies that result in globally optimal fitness. In this example, the objective function would represent the net energetic yield afforded by a specific strategy. Under this framework, MCMC simulates how ``animals" sample information by executing the task and evaluating the energy afforded by different behavioral strategies (i.e., parameter values). Behavioral strategies that consistently produce less favorable objective function values are less likely to accumulate probability mass in the posteriors.

One complete run of the MCMC algorithm, which we henceforth refer to as a ``chain", consists of many iterations. In each iteration the sampler draws random parameter values and calls the objective function at those values, either accepting or rejecting the parameters based on the function value. The number of iterations in a chain has important mathematical and biological interpretations. Chains with more iterations allow for more extensive modification of the priors, which biologically represent a simulated animal's relative belief in different behavioral strategies. With that in mind, we suggest that the number of iterations in a chain represents the amount of information the animal gathers in its environment. We can more effectively ensure that the animal consistently develops the same posterior belief in identically parameterized, but randomly independent, chains when these chains have more iterations (this is mathematically akin to ensuring the algorithm converges; \citealp{Raftery1992, Cowles1996}). Some MCMC algorithms leave iterations at the beginning of the chain out of the posterior distribution, classifying them as ``burn-in" iterations. The burn-in period was designed to enhance chain convergence \citep{Cowles1996} but by omitting the behavioral strategies employed at the beginning of the simulation process, the posterior distributions no longer include information the animal gathered during the unrealistically ``naive" (given the structure of the priors) stages of learning.

During the sampling process, MCMC allows for the acceptance of suboptimal objective function values (i.e., lower than previous values) to search the parameter space more completely and avoid local optima. The rate at which these suboptimal values are accepted can be likened to the range of behavioral strategies an animal may try in a given environment. Animals that accept a wide variety of strategies, even when they may not be optimal, could be thought of as displaying behavioral plasticity. Consistently following the optimal behavioral strategy could be thought of as displaying environmental canalization, a term used to characterize a lack of phenotypic variation in reaction to environmental change \citep{Gibson2000, Gaillard2003, Liefting2009}. The simplest way to enforce this in the model is to introduce an exponent $k > 0$ which is applied to the objective function during sampling. We can think of $k$ as an index of canalization, implying that lower values of $k$ correspond to high behavioral plasticity. Animals that possess high plasticity frequently sample many behavioral strategies amid environmental uncertainty in what is commonly referred to as bet-hedging \citep{Donaldson2008, Nevoux2010}.

We expanded on existing implementations of SDT by coupling an individual-based simulation model for animal movement with memory \citep{Avgar2013} to a Bayesian model simulating how animals learn to forage optimally. Our algorithm incorporates an objective function measuring the net energetic intake of a foraging bout, given a set of parameters controlling animal behavior. To this end, the posterior distribution of these parameters obtained after sampling reflects what simulated animals learned about the relative optimality of different foraging techniques. We tested how effectively animals adjusted to unexpected and abrupt changes in the distribution and abundance of resources on the landscape. We found that animals with higher behavioral plasticity performed more efficient foraging returns after these abrupt changes, but were less efficient when the environment did not change. Our framework displays how SDT can be extended to the simulation of realistic ecological processes that, if formulated correctly, can make effective predictions when data are lacking.

\section*{Methods}

\subsection*{The learning model}

We used Bayesian Markov Chain Monte Carlo (MCMC) sampling to simulate how animals learn to adjust their behavior based on indicators of success. The effectiveness with which an animal executes a certain task was quantified by an objective function $f$. Animals ``sample" different parameter values (i.e., behavioral strategies) and evaluate their optimality by calculating $f$; depending on the value of $f$, the animal may be more or less likely to attempt similar strategies as represented by the posterior distribution of behavioral strategies.

\begin{table}[p]
\centering
{\renewcommand{\arraystretch}{1.2}
\begin{tabular}{|c|c|c|}
\hline
\textbf{Par} & \textbf{Description} & \textbf{Value} \\
\hline
\multicolumn{3}{|c|}{\textbf{MCMC algorithm parameters}} \\
\hline
$N_{iter}$ & Number of MCMC iterations per chain & 2000 \\
$N_{burn}$ & Number of iterations in burn-in period & 500 \\
$k$ & Exponent of objective function $f$ & Many values \\
\hline
\multicolumn{3}{|c|}{\textbf{Behavioral parameters}} \\
\hline
$\beta$ & Degree of reliance on memory & Not fixed \\
$\gamma$ & Likelihood to make long navigations & Not fixed \\
$q$ & Default expectation of habitat quality & Not fixed \\
$h$ & Relative preference for resource $Q_1$ & Not fixed \\
\hline
\multicolumn{3}{|c|}{\textbf{Movement parameters} (see Appendix A)} \\
\hline
$N_r$ & Number of potential points of interest simulated & 1000 \\
$\lambda$ & Exponent of $C$ values when choosing point of interest & 10 \\
$\rho$ & Average step length on navigations & 2 \\
$\kappa$ & von Mises angular correlation parameter for navigations & 10 \\
\hline
\multicolumn{3}{|c|}{\textbf{Objective function parameters}} \\
\hline
$T_{train}$ & Length of training portion of each track & 1000 \\
$T_{test}$ & Length of test portion of each track & 1000 \\
$v$ & Energetic loss per 1 cell length of movement & 0.05 \\
$N_{avg}$ & Number of tracks incorporated into one $f$ call & 5 \\
\hline
\multicolumn{3}{|c|}{\textbf{Landscape parameters} (see Appendix A)} \\
\hline
$\underline{Q}$ & Threshold for landscape patches & 0.6 or 0.9 \\
$d_L$ & Rate of resource depletion per time step & 1.0 \\
$r_L$ & Recovery rate of depleted resources per time step & 0.025 \\
\hline
\end{tabular}}
\caption{Description of model parameters. The four parameters under the section ``Behavioral parameters" are incorporated into the objective function $f$, and sampled in the Bayesian MCMC algorithm.}
\label{tab:pardescriptions}
\end{table}

We parameterized the MCMC sampler in a way that produced consistent and biologically realistic results. We used uniform priors for each of the behavioral parameters under the assumption that animals were not initially biased towards any strategy. Using relatively uninformative priors necessitated that we added a burn-in period to our chains, and we chose to omit the first $N_{burn} = 500$ iterations of each chain from the posterior distribution to this end. Choosing the number of iterations per chain (including burn-in), $N_{iter}$, was a careful optimization of the trade-off between computational expense and consistency. Chains with more iterations take longer to simulate but they also more accurately represent what simulated animals have learned. We analyzed chains of different sizes to evaluate the fewest iterations necessary to produce consistent posterior distributions, finding that $N_{iter} = 2000$ optimized the trade-off (see Appendix B for more detail). This produced posterior distributions with $N_{iter} - N_{burn} = 1500$ parameter values. We also tested many different values for $k$, the exponent applied to $f$ during sampling: $k = 5, 10, 20, 50, 100, 200, 500, 1000$. Parameter values used in this study are summarized in Table \ref{tab:pardescriptions}. We ran our algorithm in Julia 1.6.2 using the Turing library, which offers a number of different MCMC samplers. We used a static sampler that did not require tuning or calculating a proposal function, which typically requires a function gradient \citep{Sengupta2015}. Parameters with infinite support were log-transformed and bounded on finite intervals determined by assessing their biological meaning.

\subsection*{Application of the model to foraging}

We tested our modelling framework with an optimal foraging task involving the individual-based simulation of animal movement across a continuous-space landscape. Our individual-based model (IBM) for movement is heavily inspired by \cite{Avgar2013} and contains four parameters mediating the behavioral strategy of simulated animals. We provide a summary of the model and parameters below, but see Appendix A for a more detailed explanation of the process using the ODD (Overview, Design Concepts, and Details) protocol \citep{Grimm2006b}.

Simulated animals move on a landscape characterized as a a 100 x 100 arbitrary length unit (lu) square in two-dimensional continuous space. The landscape has two independently distributed ``resources" that provide an energetic benefit to the animal. In the interest of producing movements similar to empirically observed location data, animals take discrete-time ``steps" every 1 aribtrary time unit (tu). Animals perceive, remember, and recall the quality of previously visited foraging patches to make informed movement decisions. We make four key assumptions about how animals do this, listed below:

\begin{itemize}
    \item[(A1)] Animals exhibit a preference for one of the two resources on the landscape and bias their movements accordingly.
    \item[(A2)] Animals remember the resource density of areas they have previously visited, but the animals' reliance on memory decreases over time.
    \item[(A3)] All points that the animal has not visited are perceived by the animal as having equal value, regardless of their spatial or temporal position.
    \item[(A4)] Animals are more likely to navigate to nearby points, all else held equal.
\end{itemize}

\begin{figure}
    \centering
    \includegraphics[width=250pt]{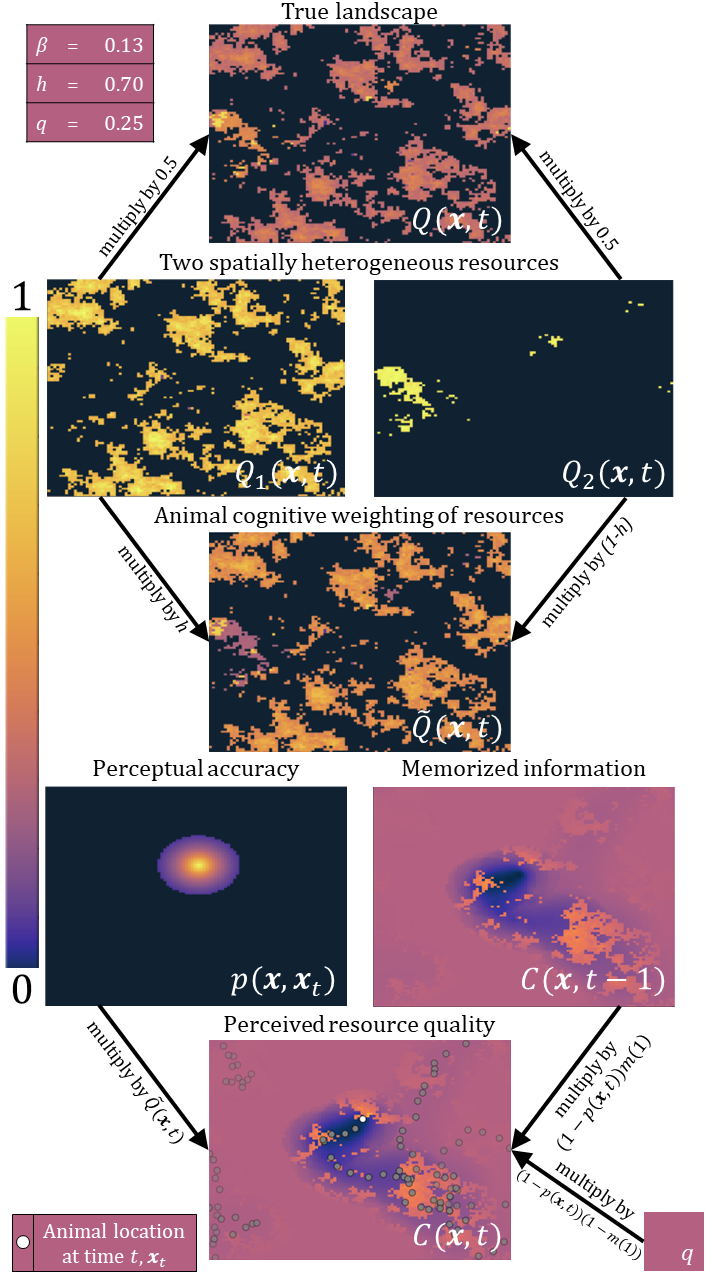}
    \caption{Schematic describing the generation of $C(\mathbf{x},t)$, the animal’s estimation of resource quality across the environment. The animal weights two independently distributed resources and incorporates newly perceived information into $C$ based on the perception function $p(\mathbf{x},\mathbf{x}_t)$. Note the incremental updating of $C$ as the animal moves to a new location ($\mathbf{x}_t$, pictured in blue on the bottom right).}
    \label{fig:cogmap}
\end{figure}

The foraging quality of any point $\mathbf{x}$ at any time $t$, which we denote $Q(\mathbf{x}, t)$, ranges from 0 to 1 and is composed of two independent foraging resources, $Q_1(\mathbf{x}, t)$ and $Q_2(\mathbf{x}, t)$. While $Q(\mathbf{x}, t) = (Q_1(\mathbf{x}, t) + Q_2(\mathbf{x}, t)) / 2$ across the landscape, we allow animals to exhibit ``habitat selection" for the different resources on the landscape (Assumption A1). The behavioral parameter $h$ ranges from 0 to 1 and mediates the animal's relative preference for $Q_1$ and $Q_2$. Simulated animals perceive $Q_1$ and $Q_2$ as independent entities, and when computing the animal's perceived foraging quality for any point $\mathbf{x}$ and time $t$, we use $\tilde{Q}(\mathbf{x},t) = hQ_1(\mathbf{x},t) + (1-h)Q_2(\mathbf{x},t)$ as opposed to $Q(\mathbf{x}, t)$ (Figure \ref{fig:cogmap}).

Simulated animals perceive new information about resources on the landscape and encode this information into spatial memory. Many different animals use memory to guide their foraging movements \citep{Panakhova1984, Clayton1998, Schlagel2014, Potts2016, Bracis2018, Ranc2021}, but heavy reliance on spatial memory is accompanied by numerous energetic costs \citep{Fagan2013}. The behavioral parameter $\beta \geq 0$ quantifies the extent to which simulated animals rely on their memory of previous foraging experiences. As $\beta$ increases the animal relies less on its memory, potentially a strategy to adapt to temporally variable environments \citep{Fagan2013}. We note that unlike memory decay, a neurological process \citep{Thomas1979}, the mechanism displayed here represents the animal's conscious choice not to rely on the memory of previous experiences.

Animals make a naive, uninformed ``guess" about the resource quality of locations they have not visited, and per Assumption A3, this guess is constant across space and time. Specifically, any location will be assigned the value $q \in [0,1]$ as long as that location remains unvisited by the animal. Larger values of $q$ suggest that the animal is more ``optimistic" about the quality of unexplored areas \citep{Berger2012, Avgar2013}, and will more frequently visit these areas as a result.

Once the animal generates an expectation of resource quality across the landscape, it must choose a location to navigate to. Assumption A4 states that animals are more likely to navigate towards nearby points than faraway points. This idea follows logically from the marginal value theorem \citep{Charnov1976}, which considers the energetic cost of travel to other patches. We included behavioral parameter $\gamma \geq 0$ to quantify this relationship. As $\gamma$ increases, the probability that the animal will navigate to a faraway point decreases; even if the animal believes there are resources far away, it may opt for nearby resource patches instead, a tactic many animals adopt as a risk avoidance mechanism \citep{Gehr2020}.

The animal's perceived resource quality for any point $\mathbf{x}$ and time $t$, denoted $C(\mathbf{x}, t)$, depends on these four assumptions. This function consists of a weighted average of three quantities: newly perceived information (weighted by perception function $p(\mathbf{x}, \mathbf{y})$), memorized information (weighted by memory function $m(t)$), and the naive expectation $q$.

\begin{align}\label{eq:C}
p(\mathbf{x}, \mathbf{y}) & = \exp\left(-\frac{d(\mathbf{x}, \mathbf{y})}{\rho}\right), \\
m(t) & = \exp(-\beta t), \\
C(\mathbf{x}, t) & = \underbrace{p(\mathbf{x}, \mathbf{x}_t) \tilde{Q}(\mathbf{x}, t)}_\text{perception} + \nonumber \\ & (1 - p(\mathbf{x}, \mathbf{x}_t)) \Big( \underbrace{m(1) C(\mathbf{x}, t-1)}_\text{memory} + \underbrace{(1 - m(1))q}_\text{expectation}\Big).
\end{align}

The perception function relies on the assumption that animals perceive nearby information more accurately than faraway information \citep{Fletcher2013, Avgar2015, Fagan2017}, where $d(\mathbf{x}, \mathbf{y})$ is the distance between $\mathbf{x}$ and $\mathbf{y}$ and $\rho$ is the animal's average movement speed in lu/tu. A positive association between movement capability and perceptual range has been documented across many animal taxa \citep{Kiltie2000, Moller2010}.

\subsubsection*{Calculating the objective function}

We designed an objective function $f$ measuring the energetic benefit afforded by a certain behavioral strategy. We divided these simulated foraging bouts into ``training" and ``test" sections of durations $T_{train}$ and $T_{test}$, respectively, and only measured $f$ over the test section. \cite{Avgar2013} made a similar correction to allow animals to develop an initial memory of their simulated landscape, producing movement paths that resemble empirically collected animal location data. We subtracted the animal's total resource intake across the simulation by the energetic loss as a result of movement, calculated as the animal's total distance traveled multiplied by a proportionality constant $v \geq 0$ (Table \ref{tab:pardescriptions}). Our function $f$ consists of an average of $N_{avg}$ independent movement tracks so it effectively characterizes the expected value of any parameter combination. We define $f_i$, the net energetic intake from the $i^{th}$ of these tracks, by summing the energetic gains collected at each location $\mathbf{x}_t$ along the animal's path:

\begin{equation}\label{eq:fi}
f_i(\beta, \gamma, q, h | \mathbf{Q}) = \frac{\sum_{t=T_{train}+1}^{T_{train}+T_{test}} Q(\mathbf{x}_t, t) - v\sum_{t=T_{train}+1}^{T_{train}+T_{test}} d(\mathbf{x}_t, \mathbf{x}_{t-1})}{T_{test}},
\end{equation}

\begin{equation}\label{eq:f}
f(\beta, \gamma, q, h | \mathbf{Q}) = \frac{1}{N_{avg}}\sum_{i=1}^{N_{avg}} f_i(\beta, \gamma, q, h | \mathbf{Q}).
\end{equation}

\subsubsection*{Scenarios of environmental change}

\begin{figure}
    \centering
    \includegraphics[width=\textwidth]{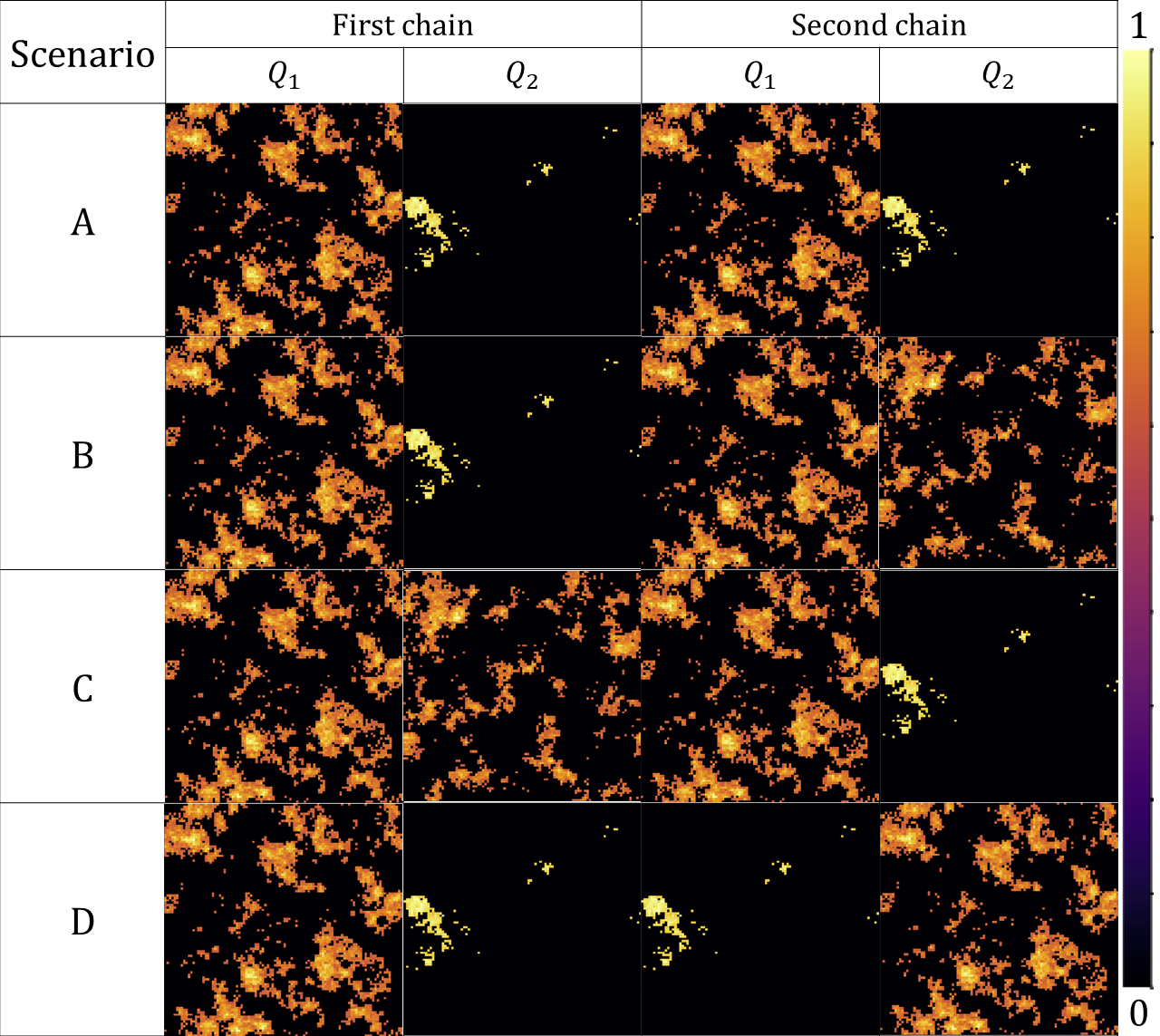}
    \caption{Different scenarios of environmental change used in our simulations. Scenario A is a “control” where the environment, composed of two resources $Q_1$ and $Q_2$, does not change at all. In Scenarios B and C, $Q_1$ stays the same, but $Q_2$ becomes more or less abundant than $Q_1$, respectively. In Scenario D, the distributions of $Q_1$ and $Q_2$ ``swap".}
    \label{fig:scenarios}
\end{figure}

We randomly generated spatially autocorrelated resource landscapes (see Appendix A for further detail) and used them to simulate abrupt landscape-level changes in the environment. Bayesian inference allows for the iterative updating of prior expectations based on previous analyses \citep{Ellison2004}. The posterior distributions of our behavioral parameters represent knowledge accumulated by a simulated animal, which we can use as more ``informative" priors for a second MCMC chain. Each of our ``scenarios" of environmental change contains two stages, where each stage has a unique $Q_1$ and $Q_2$ (Figure \ref{fig:scenarios}). The scenarios we generated incorporate two ``types" of landscape, which can be visually compared in the first chain of Scenario A (Figure \ref{fig:scenarios}). Here, $Q_1$ is much more abundant and widely distributed than $Q_2$, but $Q_2$ is richer than $Q_1$ in the small area where it can be found. Scenario A serves as a ``control" where the environment does not change; we would expect the animal to identify an optimal strategy and retain this strategy for both chains. In Scenarios B and C, only one of the resources switches between being widely abundant and locally available (the difference being the directionality of this change), and in Scenario D, both resources swap.

We ran the MCMC algorithm with each of the four scenarios and a suite of $k$ values (5, 10, 20, 50, 100, 200, 500, 1000) to evaluate how these quantities affected optimal foraging behavior. For each value of $k$ and scenario, we ran algorithm 12 independent times. We obtained posteriors for the first and second chains of each run for the four parameters $\beta, \gamma, h$, and $q$, along with a posterior distribution of $f_i$ values (1500 iterations after burn-in $\times$ 5 $f_i$ per $f$ call $\times$ 12 chains = 90000 total $f_i$ calls) for each $k$ and scenario.

\section*{Results}

\subsection*{Posterior distribution of parameters}

\begin{figure}
    \centering
    \includegraphics[width=250pt]{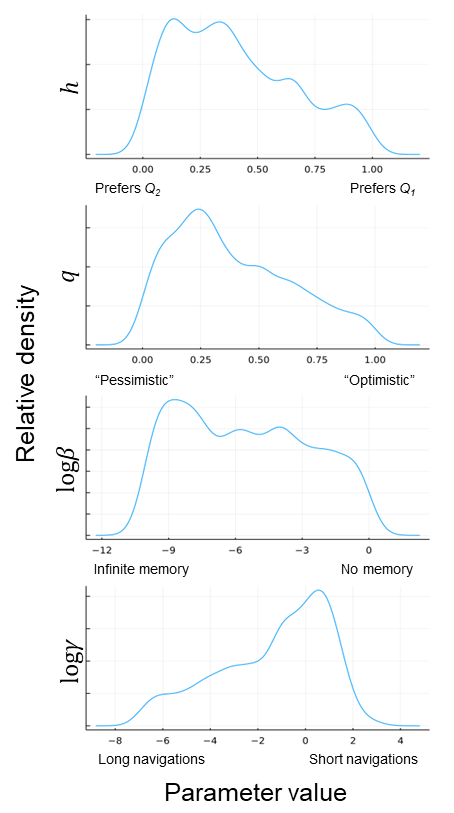}
    \caption{Posterior density plots for one independent runs of the MCMC algorithm, taken from the first chain of Scenario A (see Figure \ref{fig:scenarios}) with $k = 10$. Greater probability mass at certain parameter values indicates higher belief in that value optimizing the net energetic gain function $f$. Note, in particular, the animal’s preference for resource $Q_2$, which in this case is much less widely available but provides a larger energetic benefit  than $Q_1$ where it can be found (Figure \ref{fig:scenarios}).}
    \label{fig:density}
\end{figure}

Under the same environmental circumstances, 12 independently simulated MCMC runs produced similar posterior distributions, suggesting that $N_{iter} = 2000$ and $N_{burn} = 500$ is sufficient for convergence (a subset of these are displayed in Figure \ref{fig:density}). In most circumstances, simulated animals displayed a relatively ``pessimistic" expectation of unvisited food patches, as suggested by posterior distributions concentrated around low values of $q$. Posterior distributions of $\beta$ were relatively spread out across all values, suggesting that long-term reliance on memory only has a minimal advantage over short-term reliance in these simulations. Simulated animals avoided long-distance navigations, opting instead for values of $\gamma$ close to 1 frequently (Figure \ref{fig:density}). Most notably, though, animals simulated in Scenario A (Figure \ref{fig:scenarios}) exhibited a strong preference for resource $Q_2$, which was much less abundant across the landscape than $Q_1$. This is indicated by posterior distributions for $h$ concentrated around lower values.

\subsection*{Posterior distribution of objective function values}

\begin{figure}
    \centering
    \includegraphics[width=150pt]{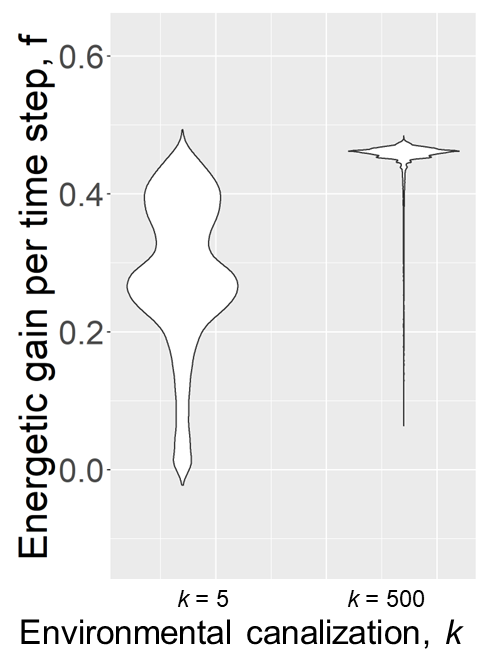}
    \caption{Example violinplots detailing the distribution of objective function $f_i$, which represents the net energetic gain from a simulated animal foraging bout. These two violinplots are taken from the second chain of Scenario A (see Figure \ref{fig:scenarios}), with $k$ taking on two different values.}
    \label{fig:violink}
\end{figure}

\begin{figure}
    \centering
    \includegraphics[width=150pt]{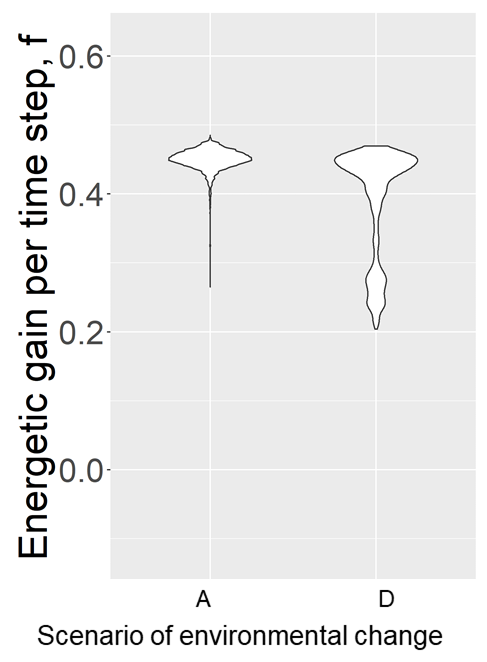}
    \caption{Example violinplots detailing the distribution of objective function $f_i$, which represents the net energetic gain from a simulated animal foraging bout. These two violinplots are drawn from the second chain of Scenarios A and D (Figure 2), respectively, with $k = 200$ for each.}
    \label{fig:violinscen}
\end{figure}

Both the scenario of environmental change and the MCMC parameter $k$ affected the second chain's posterior distribution of $f_i$ values. Typically, the spread of these distributions increased as $k$ decreased, especially in Scenario A, where they appear similar to delta functions at $k = 500$ and $k = 1000$ (Figure \ref{fig:violink}). In scenarios where the environment changed dramatically (e.g., Scenario D; Figure \ref{fig:scenarios}), these distributions took on different shapes, sometimes becoming bimodal (Figure \ref{fig:violinscen}).

\begin{figure}
    \centering
    \includegraphics[width=\textwidth]{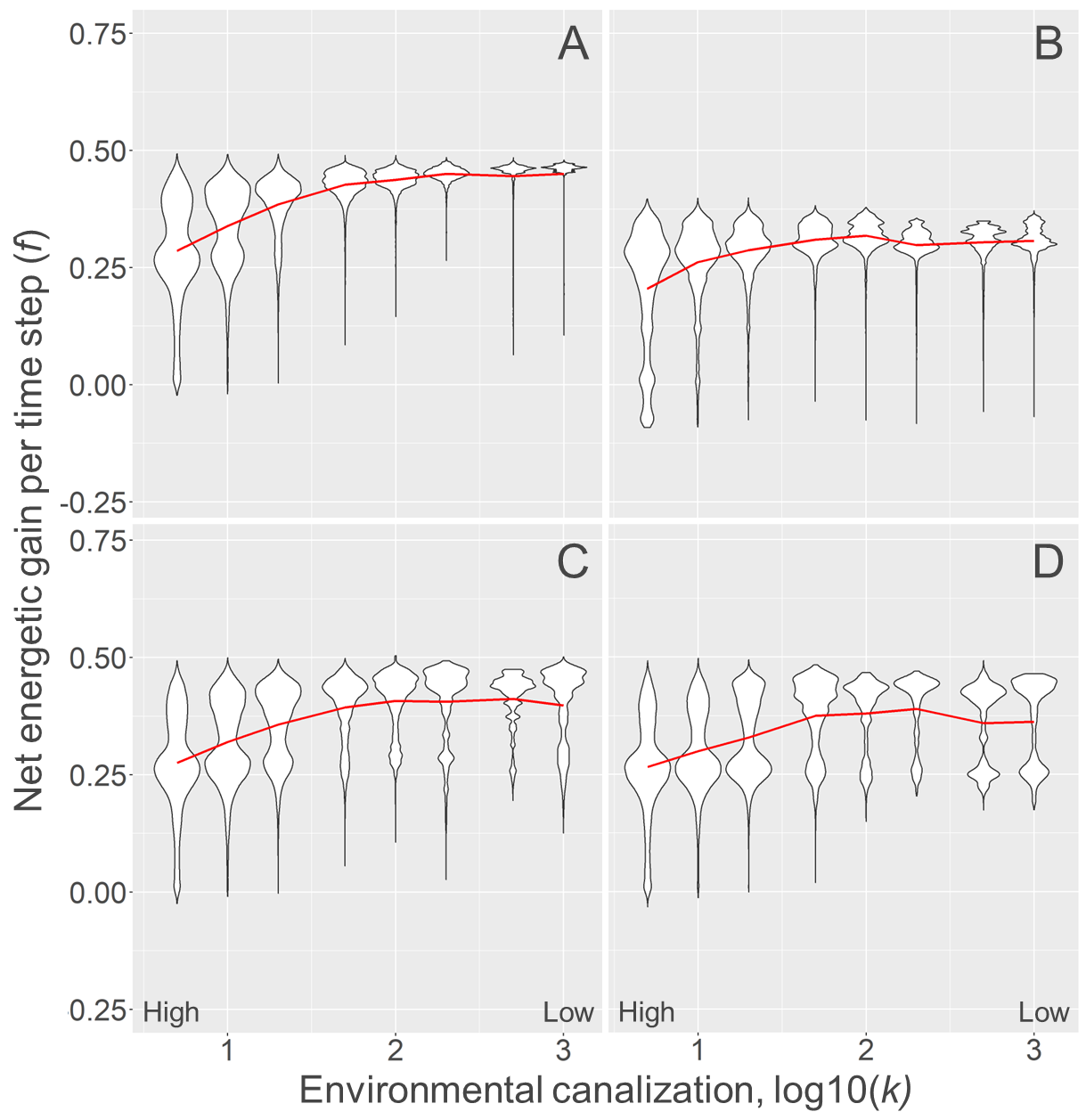}
    \caption{Effect of MCMC parameter $k$ on foraging efficiency in simulated animals under four different scenarios of temporal environmental change (see Figure \ref{fig:scenarios} for detail on each scenario). Each individual violinplot represents a sample of 90000 $f_i$ values (12 independent runs of MCMC $\times$ 1500 $f$ values per run $\times$ 5 $f_i$ values per $f$ call) representing the net energetic gain from a single simulated movement track. The red line represents the mean of all $f_i$ values from each $k$ value.}
    \label{fig:chains}
\end{figure}

More specifically, the effect of MCMC parameter $k$ on the distribution of objective function $f_i$ values depended on the scenario of environmental change (Figure \ref{fig:chains}). In Scenario A (Figure \ref{fig:scenarios}), simulated animals performed more consistently and efficiently with large values of $k$ than with small $k$. In Scenario B, $k$ had a much smaller effect on foraging success than Scenario A, although the spread of $f_i$ values was larger with smaller $k$ (Figure \ref{fig:chains}). The posterior distributions of $f_i$ from Scenario C resemble those from Scenario A at low $k$, but appear to take on a skewed, slightly bimodal shape at higher $k$. In Scenario D, intermediate values of $k$ ($k = 100$ and $k = 200$) produced foraging bouts that were, on average, more efficient than at large values of $k$ (Figure \ref{fig:chains}). The distribution of $f_i$ values was distinctly bimodal with large $k$, and as $k$ increased, more probability mass was concentrated in the second, lower mode.

\section*{Discussion}

Predicting how animals will adjust to environmental change is an important but complex ecological problem. We developed a Bayesian model that simulates how animals sample information about their environments to develop a posterior distribution of optimal foraging behavior. Our model builds on statistical decision theory, which has long been used to explain how animals learn from a Bayesian perspective \citep{McNamara1980, Berger1985, Dall2005}. We applied our learning model to a complex, continuous-space foraging task to be completed by simulated spatially informed foragers \citep{Avgar2013}. In the presence of two independently distributed resources with equal energetic return, animals simulated in our model prioritized resources that were concentrated within small, sparsely distributed patches. Animals that exhibited canalized behaviour displayed consistently efficient foraging returns in temporally predictable environments, but environmental canalization became maladaptive when we introduced sudden, unpredictable changes to the landscape. Our results suggest that Bayesian MCMC can be used to simulate how animals, and potentially even humans, learn a wide variety of tasks in an ever-changing world.

When faced with the choice of two resources, simulated animals chose the resource that was available in smaller, but more heavily concentrated patches (Figure \ref{fig:density}). We expected to observe a preference for the more abundant $Q_1$, but our model showed that they instead opted for the less abundant but richer $Q_2$. This finding suggests that simulated animals occupy areas with the highest possible habitat suitability, a key principle of ideal free distribution (IFD) theory \citep{Fretwell1969, Cantrell2007}. Many patterns predicted by IFD theory can be seen in our results even though our IBM did not incorporate competition between individuals (this could be an interesting topic for future work). Specifically, IFD theory predicts that individuals residing in poor habitat will adjust for the lack of resource abundance by adopting larger home ranges \citep{Hache2013}. Simulated animals in our model centralized their movements around small plentiful resource patches, producing smaller home range sizes than individuals that foraged on less concentrated resources. Similarly, the resource dispersion hypothesis predicts that animals will occupy larger home ranges when resources are less spatially concentrated \citep{Macdonald1983, Macdonald2015}. Increasing the speed or breadth of resource depletion or further decreasing the spatial availability of these concentrated resources could modify this relationship.

The wide variety of behavioral strategies adopted by simulated animals with high behavioral plasticity during sampling produced variable energetic outcomes. Behavioral plasticity allows animals to exhibit a variety of foraging strategies simply as a result of learning and adjusting to new environmental drivers \citep{Parrish2000}. Animals with highly canalized behavior (i.e., low plasticity) would be expected to perform one foraging strategy consistently \citep{Gaillard2003, Snell2013, Wong2015}, and we frequently saw that in our simulations. This is also unsurprising from an analytical perspective, since $k$ (specificity) also resembles the number of ``clones" used in data cloning algorithms \citep{Lele2007}. This consistency also suggests that there is minimal stochastic variation in the value returned by our objective function $f_i$ when our behavioral parameters were held constant.

Our simulations strongly suggest that behavioral plasticity is adaptive when the environment changes dramatically and unexpectedly. Animals simulated in temporally constant environments had unimodal distributions of energetic return, but those simulated in temporally unpredictable environments had a second mode centered around a lower energetic intake (Figure \ref{fig:violinscen}). The latter group of simulated animals foraged efficiently until the distribution of resources suddenly changed, rendering the original strategy suboptimal. Animals with high behavioral plasticity shifted their resource preferences depending on the environment, for better or for worse \citep{Parrish2000, Van2018, Dunn2020}. Animals with very low behavioral plasticity continued to forage according to their initial, now suboptimal, strategy, while animals with intermediate levels of behavioral plasticity adjusted their foraging strategies more effectively (Figure \ref{fig:chains}). Animals with very high behavioral plasticity performed a diverse array of foraging strategies, many of which were too inefficient to produce optimal foraging returns. 

While behavioral plasticity is typically considered an adaptive trait, some animals suffer from it. Ecological traps are resources that appear beneficial to animals but, in reality, do not confer a fitness benefit (e.g., mayflies lay their eggs on asphalt because it reflects light similarly to water; \citealp{Kriska1998}). Ecological traps have become more frequent in the Anthropocene due to the proliferation of man-made novel objects in natural environments \citep{Robertson2016}. A typical consequence of behavioral plasticity is an increased likelihood to explore unfamiliar stimuli \citep{Mettke2009, Snell2013}, which is believed to associate behavioral plasticity and vulnerability to ecological traps \citep{Robertson2016}. The results from our simulation study corroborate empirical evidence that environmental canalization can be more effective than behavioral plasticity in some environments.

Translocated animals represent an effective way to test our model, displaying behavior similar to our simulations. Animal translocation and reintroduction protocols have many purposes, ranging from the displacement of potentially dangerous animals \citep{Milligan2018} to the restoration of populations and ecosystems \citep{Seddon2007, Polak2011}. Translocated animals are abruptly brought to entirely new environments where they must learn to forage optimally or face heightened mortality risk. The nature of these protocols makes them an effective real-life test for our model, and many of the predictions offered by our model are verified from translocation studies. Translocated elk (\textit{Cervus canadensis}) displayed different foraging behavior depending on the environmental conditions in their original home range and the environmental change they underwent \citep{Falcon2021}. Specifically, elk translocated between two very different environments (resembling our Scenario D) were more exploratory and less reliant on memory than those translocated between similar environments, suggesting a shift in behavior from their original home ranges \citep{Falcon2021}. As another case study, greater prairie-chickens (\textit{Tympanuchus cupido}) typically sought out habitat similar to that of their natal ranges, suggesting a strong prior preference for resources found in their old environments \citep{Kemink2013}. Here, canalization was detrimental to the birds' survival, adding support to the pattern observed in panel D of Figure \ref{fig:density}. Translocations and reintroductions are frequently practiced across a wide array of animal taxa, but they are still risky and unpredictable \citep{Berger2014}. The principles drawn from our analysis provide an improved forecast for the efficacy of these protocols in different ecological systems.

\section*{Conclusion}

We developed a modelling framework that innovatively applies the principles of Bayesian statistics to animal foraging and learning. Much of what we currently know about animal learning comes from manipulative experiments conducted with captive animals \citep{Pearce2008}. Many of these studies have been critical for unearthing the mechanisms behind animal cognition, memory, and learning \citep{Pavlov1927, Rescorla1972}, but they do not replicate the conditions wild animals experience. By incorporating the prevailing mathematical theory behind animal learning, our modelling framework fills this gap. Our results with respect to continuous-space foraging align with optimal foraging theory \citep{Charnov1976}, ideal free distribution theory \citep{Fretwell1969}, and prevailing knowledge on behavioral plasticity \citep{Wong2015, Robertson2013}. With that being said, our model for learning is general enough that it need not be confined to optimal foraging. Specifically, any problem that can be characterized in the form of an objective function and a set of parameters representing behavior is tractable for our framework. This could include movement on different spatial or temporal scales, social learning, or communication. Even more thought-provoking is the potential for our modelling framework to predict how humans learn and make decisions. While the decisions made by animals can affect the individual's fitness and survival, human decisions have the potential to reverberate much more widely, which has become even more apparent during the COVID-19 pandemic \citep{Bavel2020}. Through these potential applications and more, our computational modelling framework has the capacity to address challenging problems in cognitive science.


\section*{Author contributions}

PRT, MKL, and MAL conceived the idea for the model. PRT wrote all computer programs and ran all analyses. PRT wrote the first draft of the manuscript, upon which MKL and MAL provided extensive feedback.



\newpage{}

\section*{Appendix A: ODD Protocol}

%
%
%
%
%

\renewcommand{\theequation}{A\arabic{equation}}
\renewcommand{\thetable}{A\arabic{table}}
\renewcommand{\thefigure}{A\arabic{table}}
\setcounter{equation}{0}  
\setcounter{figure}{1}
\setcounter{table}{0}

\subsection*{Purpose}

We developed a model heavily influenced by \cite{Avgar2013} to simulate the movement of spatially informed foragers. The model includes four parameters that, when combined, quantify an animal's foraging strategy. These parameters are intended to measure behaviorally plastic qualities of an animal as opposed to genetic or morphological traits. We assessed the adaptive value of different foraging strategies using a net energetic gain metric, which weighs the animal's resource intake against the energetic cost of movement. We do not specifically liken the model to any animal taxon, but we note that many common behavioral processes (e.g., migration and sociality) are not included in the model.

\subsection*{State variables and scales}

The model consists of one individual (henceforth referred to as an ``animal") that moves throughout a bounded spatial landscape. The animal performs discrete-time, continuous-space movements at constant temporal intervals of 1 arbitrary time unit (tu). The landscape is a 100 x 100 arbitrary length unit (lu) square in two-dimensional space. Each spatial point on the landscape $\mathbf{x}$ and time index $t$ has a resource quality $Q(\mathbf{x}, t) \in [0,1]$ representing the energetic value of resources at that point. For mathematical convenience, we formulated $Q(\mathbf{x}, t)$ as a piecewise constant function; all $\mathbf{x}$ in any 1x1 lu ``grid cell" have the same value of $Q(\mathbf{x}, t)$ at any time $t$. To prevent animals from getting ``trapped" in corners or boundaries of the landscape, we assume that landscapes have wrap-around boundaries (e.g., if the animal moves far enough to the left, it will eventually end up on the right side of the grid).

The landscape has two unique resources that are added together to produce the total resource quality $Q(\mathbf{x}, t)$ for each point and time. We define $Q_1(\mathbf{x}, t)$ and $Q_2(\mathbf{x}, t)$ to be the quality values for the first and second resources at point $\mathbf{x}$ and time $t$, respectively. Both of these resource functions can take on values between 0 and 1, so to ensure that $Q(\mathbf{x}, t)$ is defined properly, we set $Q(\mathbf{x}, t) = (Q_1(\mathbf{x}, t) + Q_2(\mathbf{x}, t)) * 0.5$ for every point $\mathbf{x}$ and time $t$.

We incorporated depletion-recovery dynamics to the landscape to ensure animals would be incentivized to move. When the animal visits any point in a grid cell, it consumes and depletes that cell's resources. Specifically, we decrement $Q_1(\mathbf{x}, t)$ and $Q_2(\mathbf{x}, t)$ by resource depletion parameter $d_L$ for every point $\mathbf{x}$ in the cell the animal visits at time $t$. If $d_L$ is greater than the resource value at that time, the cell is depleted entirely and is assigned a resource value of 0. Each depleted resource recovers by $r_L$ units each time step until reaches its original, pre-depletion value. We fixed $d_L$ and $r_L$ for all simulations (Table \ref{tab:pardescriptions}).

\subsection*{Process overview and scheduling}

We tracked information storage in simulated animals using $C(\mathbf{x}, t)$, which represents the animal's estimation of resource quality for each point and time. As the animal perceives and remembers new information through movement, $C$ is updated. The animal moves by choosing a ``point of interest" to navigate to based on $C$. Points of interest may take more than 1 tu to reach, reflecting the numerous timescales at which animals make movement decisions \citep{McClintock2014, Blackwell2016}.

\subsection*{Design concepts}

\noindent \textit{Fitness:} Simulated animals perform the most basic version of ``fitness-seeking" in that they search for points with a higher concentration of resources. Following Assumption A1, animals exhibit ``habitat selection" for the different resources on the landscape. We introduce the parameter $h \in [0,1]$ to quantify this relationship. When the animal visits a new location, it stores the value of that location as $\tilde{Q}(\mathbf{x},t) = hQ_1(\mathbf{x},t) + (1-h)Q_2(\mathbf{x},t)$ rather than $Q(\mathbf{x},t)$ (Figure \ref{fig:cogmap}). Per Assumption A4, animals will be more likely to navigate to nearby points, as this minimizes locomotive cost as well as the opportunity cost of navigating through potentially resource-poor habitat on the way to a faraway point of interest. \newline

\noindent \textit{Sensing:} Animals are not omniscient and must obtain information via perception. Typically, animals perceive nearby information more accurately \citep{Avgar2015, Fagan2017}. Mathematically, we formalize this using a perception function $p(\mathbf{x}, \mathbf{y})$. This function measures how accurately (ranging from 0 to 1) an animal located at $\mathbf{x}$ perceives information about $\mathbf{y}$. We chose an exponential decay function (similar to \citealp{Avgar2013}) to represent this relationship:

\begin{equation}\label{eq:perception}
p(\mathbf{x}, \mathbf{y}) = \exp\left(-\frac{d(\mathbf{x}, \mathbf{y})}{\rho}\right),
\end{equation}

\noindent where $d(\mathbf{x}, \mathbf{y})$ is the distance between $\mathbf{x}$ and $\mathbf{y}$, accounting for wrap-around boundaries. We assume that the animal's perceptual ability increases with $\rho$, the parameter governing the animal's locomotive capability.

\noindent \textit{Memory:} Assumption A2 states that the animal's reliance on memory decreases as the time since the formation of that memory increases. Mathematically, we used an exponential decay function to represent this (similarly to \citealp{Avgar2013}). The function $m(t)$ ranges between 0 and 1 and quantifies the animal's reliance on memory as a function of how long ago the memory was formed. Simply put, $m(t) = \exp\left(-\beta t\right)$. If $\beta = 0$, the animal effectively has an infinite memory, and as $\beta$ becomes infinitely large, the animal begins to neglect its memory entirely. \newline
 
\noindent \textit{Prediction:} The animal estimates the resource quality at any point on the landscape using perception and memory, but if it has never visited a location on the grid, it must still make a naive ``guess" about the resource quality there \citep{Berger2012, Avgar2013}. Assumption A3 states that this guess is constant across space and time; in other words, the animal will treat all unvisited points equally throughout the simulation. We can represent this guess with $q \in [0,1]$. Larger values of $q$ will result in more exploratory movement as animals assign higher value to unvisited areas. \newline

\begin{figure}
    \centering
    \includegraphics[width=\textwidth]{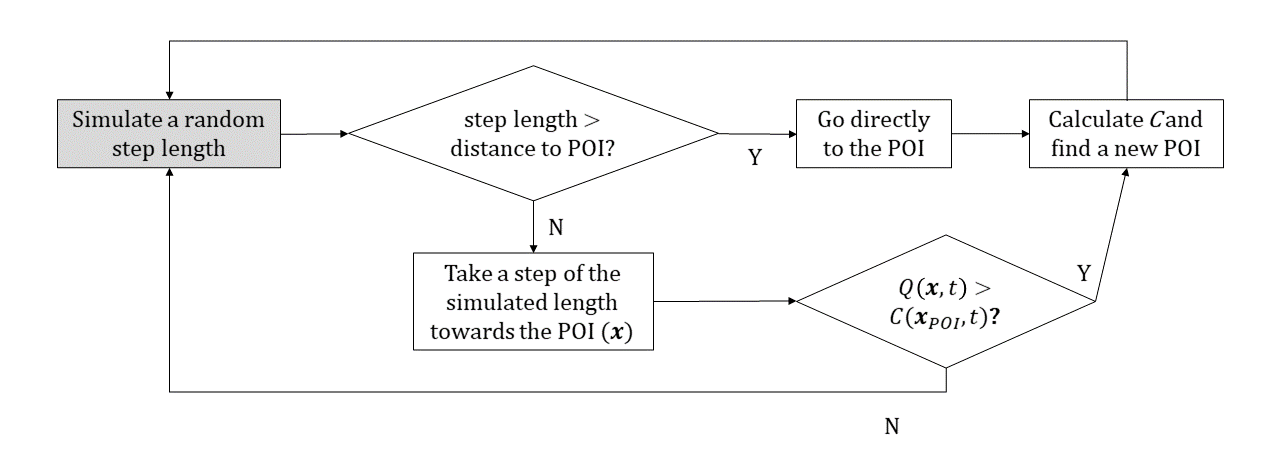}
    \caption{Flowchart describing the individual-based simulation model for animal movement. At each time step, animals update their perception of the environment $C$, occasionally using it to choose a point of interest (POI) to navigate to. This navigation can take any number of time steps, as the animal does not typically stop navigating until it reaches the point.}
    \label{fig:flowchart}
\end{figure}

\noindent \textit{Stochasticity:} Animal movement paths are stochastic, and as a result animals will not always visit the patch that confers the highest expected benefit (i.e., the highest value of $C$). That being said, points with higher values of $C$ are still more likely to be chosen as points of interest. When the animal is not currently en route to a point of interest, a new point of interest is picked using a Monte Carlo sampling technique. This involves simulating $N_r = 1000$ possible points of interest $\mathbf{x}_{t,1}, \mathbf{x}_{t,2}, ..., \mathbf{x}_{t,N_r}$ and randomly picking one (denoted $\mathbf{x}^P_t$) based on the value of $C$. More specifically,

\begin{equation}
    P(\mathbf{x}_{t, i} = \mathbf{x}^P_t) = \frac{C(\mathbf{x}_{t,i}, t)^\lambda}{\sum_{j=1}^{N_r} C(\mathbf{x}_{t,j}, t)^\lambda},
\end{equation}

\noindent for any positive integer $i \leq N_r$. We include a fixed constant $\lambda \geq 0$ that controls the ``determinism" of the animal's movements: as $\lambda$ increases, it is more likely to choose the point with the highest value of $C$.

We simulate the $\mathbf{x}_{t,i}$ as end points of a movement ``step" beginning at $\mathbf{x}_{t-1}$, where the lengths of each step follow an exponential distribution. The shape of this distribution results in smaller step lengths being more frequently sampled, following Assumption A4. We define $\gamma \geq 0$ as the ``rate" parameter of the exponential distribution, quantifying the strength of the relationship between distance and point-of-interest selection. As $\gamma$ approaches 0, every point on the grid has an equal chance of being selected (assuming equal values of $C$). If $\gamma$ is large, all $\mathbf{x}_{t,i}$ will be close to the animal and it will not undertake long-distance navigations very often.

The animal navigates to points of interest by performing a biased random walk (Figure \ref{fig:flowchart}). The lengths of each step along the navigation are simulated from a gamma distribution with mean and variance $\rho$. This distribution has an entirely positive support and is roughly bell-shaped for most values of $\rho$, including the value we used (Table \ref{tab:pardescriptions}). If simulated step lengths are longer than the distance to the point of interest (i.e., the animal would ``overshoot" its destination), the animal goes directly to the point of interest instead. Otherwise, it takes a step of the simulated length towards the point of interest. The heading of this step is simulated from a von Mises distribution where the mean heading is the heading required to reach the point of interest. The concentration parameter for this distribution, $\kappa \geq 0$, is a fixed quantity in this model (Table \ref{tab:pardescriptions}). It is recommended that large values of $\kappa$, which cause more directed movement to the point of interest, are used here. If one of the steps on the animal's navigation ends on a point that has better resources than the point of interest (i.e., $\tilde{Q}(\mathbf{x}_{t+1}, t+1) > C(\mathbf{x}^P_t, t+1)$), the animal ``forgets" about the point of interest and prioritizes foraging at the newfound location. The algorithm restarts whenever the animal arrives at its point of interest. \newline

\noindent \textit{Observation:} We collected information about the animal's movement as well as its cumulative resource intake. We keep track of the animal's location $\mathbf{x}_t$, as well as the value of $Q(\mathbf{x}_t, t)$, for each time step $t$ in the track. Note that while the animal exhibits relative preference for resources using $\tilde{Q}$, it still takes in equal amounts of both resources when it visits a patch. \newline

\noindent Our model does not implement interaction or collectives since animals are solitary on the landscape. While we assume that animals can ``adapt" to environmental conditions by modifying $\beta, \gamma, q$, and $h$ between simulations, we do not allow for adaptation within a single simulation. We are not particularly focused on emergent properties such as home range formation.

\subsection*{Initialization}

At the beginning of each simulation, we randomly generate a landscape and initialize the animal at a random point on that landscape. Initially, $C(\mathbf{x}, 0) = q$ for every point $\mathbf{x}$, as the animal has no prior experience on the grid.

\subsection*{Input}

For each simulated animal movement path, we supplied two randomly generated landscapes (for $Q_1$ and $Q_2$ respectively) as inputs. We simulated our landscapes as Gaussian random fields, implying that each cell on the grid is a component of a multivariate Gaussian random variable \citep{Schlather2012}. In this case, the covariance between any two cells depends on the wrap-around distance between the two cells (closer cells have higher covariance). We then scaled the values such they all fell between 0 and 1. 

To more accurately capture the patchiness of many real-world habitats, we defined a cut-off value $\underline{Q}$ that could be used to make these landscapes more patchy. Under this rule, any grid cell with a value of $Q$ below $\underline{Q}$ would be set to 0. Increasing $\underline{Q}$ decreases the overall resource quality of the landscape and is more likely to confine the animal to specific high-quality patches. Here, we used landscapes with $\underline{Q} = 0.6$ and $\underline{Q} = 0.9$ (Figure \ref{fig:scenarios}).

\subsection*{Submodels}

Our main submodel is the calculation of $C$, the animal's spatial map of perceived resource quality. This calculation is composed of three mechanisms: perception ($p(\mathbf{x}, \mathbf{y})$), memory ($m(t)$), and default expectation ($q$). Figure \ref{fig:cogmap} displays how these quantities are combined and weighted to produce $C$. This is mathematically formalized below:

\begin{align}\label{eq:csuppinfo}
C(\mathbf{x}, t) & = \underbrace{p(\mathbf{x}, \mathbf{x}_t) \tilde{Q}(\mathbf{x}, t)}_\text{perception} + \nonumber \\ & (1 - p(\mathbf{x}, \mathbf{x}_t)) \Big( \underbrace{m(1) C(\mathbf{x}, t-1)}_\text{memory} + \underbrace{(1 - m(1))q}_\text{expectation}\Big).
\end{align}

\noindent Note that $m(1) = exp(-\beta)$, which resembles the model from \cite{Avgar2013}.

\newpage{}
\renewcommand{\theequation}{B\arabic{equation}}
\renewcommand{\thetable}{B\arabic{table}}
\renewcommand{\thefigure}{B\arabic{table}}
\setcounter{equation}{0}  
\setcounter{figure}{1}
\setcounter{table}{0}

\section*{Appendix B: Determining the appropriate number of MCMC iterations}

We determined an optimal number of iterations per MCMC chain by identifying when additional iterations did not substantially affect the posterior distribution of the four behavioral parameters. If some value $N$ were to be sufficient as the number of iterations per chain, we would expect that a chain simulated for $N$ iterations would produce similar posteriors when we added additional iterations to the chain. If simulating more iterations produced negligibly different posteriors, it is not computationally worthwhile to perform those iterations. To that end, we ran a chain of the MCMC sampling algorithm for our foraging task with 5000 iterations (what we deemed to be the largest computationally reasonable value). We then took the first $N$ iterations of that chain and compared the posterior distribution from that subset with a slightly larger subset, the first $N + 500$ iterations. We used a static MCMC sampler in our analysis so the individual iterations were independent of each other, rendering this process similar to comparing two separate chains. 

\begin{figure}
    \centering
    \includegraphics[width=\textwidth]{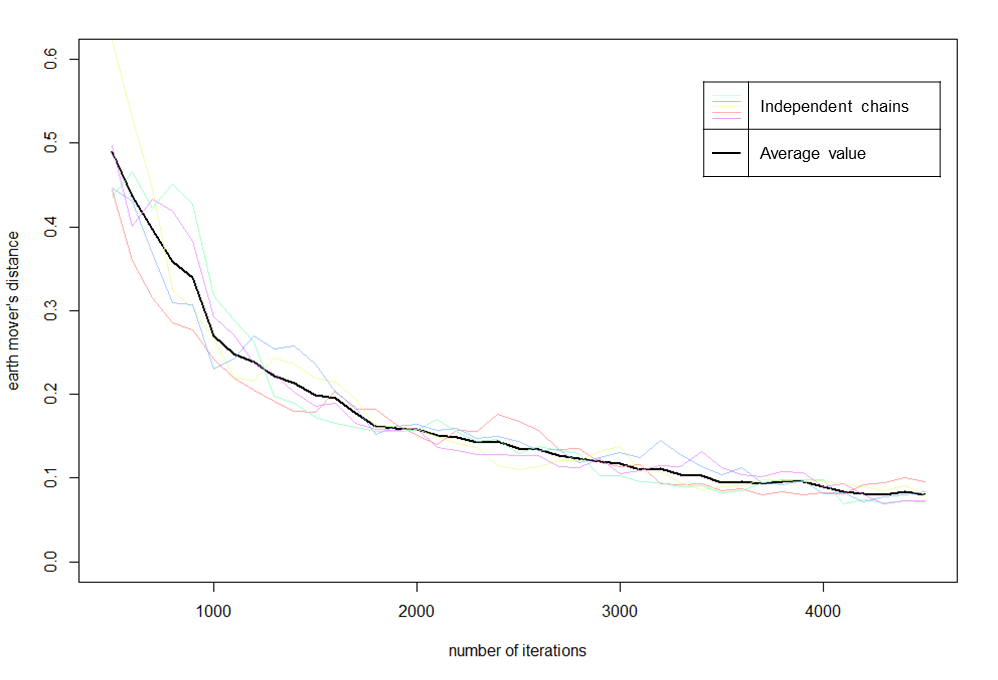}
    \caption{Relationship between the number of iterations in a MCMC chain used to simulate the foraging task and distributional similarity, measured using the earth mover's distance. We calculated the earth mover's distance between the first $N$ iterations of the chain and the first $N+500$ iterations to evaluate the difference that adding 500 iterations would make to the posterior distribution of animal behaviour. The coloured lines represent five individual runs of the process, and the thicker black line represents the mean earth mover's distance across these runs.}
    \label{fig:EMD}
\end{figure}

We compared posterior distributions using the earth mover's distance, also known as the Wasserstein distance, a common tool for comparing multivariate distributions across many fields \citep{Vaserstein1969, Rubner2000, Potts2014}. The earth mover's distance approximates the energy required to spatially transform one probability distribution such that it resembles another. As a result, lower values of this metric suggest higher distributional similarity, and an earth mover's distance of 0 is only achieved between two perfectly identical probability distributions. Plotting the earth mover's distance against $N$, the proposed number of iterations, for many different values of $N$ (ranging from 500 to 4500 by 100) led us to identify $N_{iter} = 2000$ as the appropriate number of iterations (Figure \ref{fig:EMD}). We ran the process described above five independent times to ensure that this relationship was similar with different random runs of the algorithm.

\newpage{}

%
%

\end{document}